\documentclass[pra,aps,reprint,a4paper,superscriptaddress,longbibliography,floatfix]{revtex4-1}
\usepackage{times}
\usepackage{epsfig}
\usepackage{amsfonts}
\usepackage{amsmath}
\usepackage{amssymb}
\usepackage{color}
\usepackage{multirow}
\usepackage[normalem]{ulem}
\usepackage{latexsym}
\usepackage{amsfonts}
\usepackage{mathrsfs}
\usepackage{natbib}
\usepackage{verbatim}
\usepackage[T1]{fontenc}

\usepackage[colorlinks=true,linkcolor=blue,citecolor=magenta,urlcolor=blue]{hyperref}

\DeclareMathOperator{\Tr}{tr}

\newcommand{\ket}[1]{|#1\rangle}
\newcommand{\bra}[1]{\langle#1|}
\newcommand{\braket}[2]{\langle#1|#2\rangle}

\newcommand{\lmax}[0]{\lambda_{\textnormal{max}}}

\begin{document}


\title{Self-testing quantum states and measurements in the prepare-and-measure scenario
}


\author{Armin Tavakoli}
\affiliation{D\'epartement de Physique Appliqu\'ee, Universit\'e de Gen\`eve, CH-1211 Gen\`eve, Switzerland}

\author{J\k{e}drzej Kaniewski}
\affiliation{QMATH, Department of Mathematical Sciences, University of Copenhagen,
	Universitetsparken 5, 2100 Copenhagen, Denmark}

\author{Tam\'as V\'ertesi}
\affiliation{Institute for Nuclear Research, Hungarian Academy of Sciences, P.O. Box 51, 4001 Debrecen, Hungary}

\author{Denis Rosset}
\affiliation{Perimeter Institute for Theoretical Physics, 31 Caroline St. N, Waterloo, Ontario, Canada, N2L 2Y5}
\affiliation{Institute for Quantum Optics and Quantum Information (IQOQI), Boltzmangasse 3, 1090 Vienna, Austria}

\author{Nicolas Brunner}
\affiliation{D\'epartement de Physique Appliqu\'ee, Universit\'e de Gen\`eve, CH-1211 Gen\`eve, Switzerland}


\begin{abstract}
The goal of self-testing is to characterise an a priori unknown quantum system based solely on measurement statistics, i.e.~using an uncharacterised measurement device. Here we develop self-testing methods for quantum prepare-and-measure experiments, thus not necessarily relying on entanglement and/or violation of a Bell inequality. We present noise-robust techniques for self-testing sets of quantum states and measurements, assuming an upper bound on the Hilbert space dimension. We discuss in detail the case of a $2 \rightarrow 1$ random access code with qubits, for which we provide analytically optimal self-tests. The simplicity and noise robustness of our methods should make them directly applicable to experiments. 

\end{abstract}


\maketitle


\textit{Introduction.---} Predicting the results of measurements performed on a given physical system has traditionally been the main concern of physics. However, with the advent of device-independent quantum information processing \cite{ABG07,Col07,PAM10}, the opposite question has become relevant. More specifically, given an initially unknown system and an uncharacterised measurement device, what can be inferred about the physics of the experiment based solely on the observed measurement statistics? Despite the apparent generality of this question, certain cases do allow for a precise characterization of the system. This is referred to as self-testing \cite{MY98, MY04}.  

The possibility to self-test quantum states and measurements usually relies on quantum nonlocality. Consider two distant observers performing local measurements on a shared quantum state. When the resulting statistics leads to violation of a Bell inequality \cite{bell}, it is necessarily the case that the shared quantum state is entangled, and moreover, that the local quantum measurements are incompatible; see e.g.~\cite{review}. Furthermore, for specific Bell inequalities, maximal violation (i.e.~the largest possible value in quantum theory) implies that the quantum state and the measurements can be uniquely identified (up to local isometries). For instance, a maximal violation of the Clauser-Horne-Shimony-Holt (CHSH) Bell inequality \cite{CHSH69} implies maximally incompatible measurements (two anti-commuting Pauli observables) and a shared maximally entangled two-qubit state \cite{SW87, PR92, T93,RUV}. More recently, it has been demonstrated that all bipartite pure entangled states can be self-tested \cite{CG17}, as well as certain multipartite entangled states \cite{M14,PVN14,WCY14}. Another important progress is the development of self-testing methods robust to noise \cite{BLM09,MYS12,YN13,BP15,swaptrick,Kan16,Kan17}. For instance, given a certain level of violation of a Bell inequality (but not necessarily maximal), the fidelity between the initially unknown state and a given target state can be lower-bounded. 

Self-testing thus offers promising perspectives for the certification of quantum systems in experiments (see e.g.~\cite{wineland}), as well as for device-independent quantum information protocols \cite{supic}. It is therefore natural to ask whether the concept of self-testing can be applied to more general quantum experiments, beyond those based on entanglement and nonlocality.

In the present work, we develop self-testing methods tailored to the prepare-and-measure scenario. This covers a broad class of experiments, where quantum communication schemes (e.g.~the BB84 quantum key distribution (QKD) protocol) are prominent examples. In this setting, a preparation device initially prepares a quantum system in different possible states. The system is then transmitted to a measurement device, which performs different possible measurements on it. While it is still possible in this case to characterise certain physical properties of the system based only on statistics, this requires in general an assumption on the devices. One possibility, which we will follow here, is to assume that the set of quantum states and measurements admit a full description in a Hilbert space of given dimension \cite{GB10,Hendrych,Ahrens}. Intuitively this means that the amount of information communicated from the preparation device to the measurement device is assumed to be upper bounded. Such a scenario considering quantum systems of fixed dimension, but otherwise uncharacterised, is referred to as semi-device-independent, and opens interesting possibilities for quantum information processing \cite{PB11,LYW11,WP15,Lunghi,MT16}.

Here we demonstrate techniques for robustly self-testing sets of prepared quantum states, as well as sets of quantum measurements. These methods allow one to (i) assess the compatibility of given sets of preparations and measurements with the observed statistics, and (ii) lower-bound the average fidelity between the unknown preparations (measurements) and a set of ideal quantum states (measurements). We discuss in detail a simple prepare-and-measure scenario, namely the $2 \rightarrow 1$ random access code (RAC). This allows us to provide analytically optimal self-tests for a pair of anti-commuting Pauli observables, and for a set of four qubit states corresponding to the eigenstates of two anti-commuting Pauli observables. We then generalise these results to other prepare-and-measure scenarios. The simplicity and robustness of our methods should make them directly applicable to experiments. We conclude with a number of open questions.

\textit{Scenario.---} We consider a quantum prepare-and-measure experiment. Upon receiving input $x$, a preparation device emits a physical system in a quantum state $\rho_x$. The system is then transmitted to a measurement device, which, upon receiving an input $y$, performs a quantum measurement returning an outcome $b$. Formally, the measurement is described by a set of positive operators $M_{y}^b$, that equal identity when summed over $b$. Importantly both the specific states $\rho_x$ and measurements $M_{y}^b$ are a priori unknown to the observer. The statistics of the experiment is then given by $P(b|x,y) = \Tr(\rho_{x} M_{y}^b)$.
In this setting, any possible probability distribution can be obtained, given that the prepared states $\rho_x$ can be taken in a sufficiently large Hilbert space. This is however no longer the case when we limit the Hilbert space dimension; specifically we impose that $\rho_x\in \mathcal{L}(\mathbb{C}^d)$ for some given $d< |x|$ (where $|x|$ denotes the number of possible inputs $x$). In this case, limits on the set of possible distributions can be captured via inequalities of the form
\begin{equation}
\mathcal{A} = \sum_{x,y,b} \alpha_{xyb} P(b  | x, y) \leq  Q_d,
\end{equation}
where $\alpha_{xyb}$ are real coefficients. These ``dimension witnesses'' allow one to place device-independent lower bounds on the dimension of the quantum system \cite{GB10}.

Subsequently, one can ask what the limitations are on the set of distributions $P(b|x,y)$ given that the preparations admit a classical $d$-dimensional representation, i.e.~there exists a $d$-dimensional basis such that all states $\rho_x$ are diagonal in this basis. We denote by $C_d$ the maximal value of the quantity $\mathcal{A}$ in this case. Interestingly, for well-chosen quantities $\mathcal{A}$, one finds that $C_d < Q_d$. Thus, for a given system dimension $d$
, quantum systems outperform classical ones, in the sense that certain quantum distributions cannot be reproduced classically \cite{GB10}. This quantum advantage can be viewed as the origin for the possibility of developing self-testing methods for the prepare-and-measure scenario; in analogy to Bell inequality violation being the root for self-testing entangled states. 

In the following we present robust self-testing techniques based on specific dimension witnesses $\mathcal{A}$. Based only on the value of $\mathcal{A}$, which is directly accessible from the experiment statistics, we characterise the (initially unknown) prepared states and measurements. In particular, when the maximal value of the witness is obtained, i.e.~$\mathcal{A} = Q_d$, then a specific set of pure states $\rho_x = \ket{\psi_x} \bra{\psi_x}$ and a specific set of projective measurements $M_y^b$ must have been used (up to a unitary). Moreover, when a non-maximal value $ \mathcal{A} < Q_d$ is obtained, the compatibility of given sets of preparations and measurements can be assessed. Finally, one can efficiently lower bound the fidelity between the prepared states and measurements and the ideal (or target) states and measurement leading to $\mathcal{A} = Q_d$. 

Note that a recent series of works followed a related though conceptually different approach, based on hypothesis testing \cite{michele1,michele2,michele3}. This method does however not allow for self-testing.
	 
\textit{The $2\rightarrow 1$ random access code.---} We discuss in detail a simple prepare-and-measure experiment. This involves four possible preparations, denoted by $x = (x_0, x_1)$ (where $x_j \in\{0,1\}$ ), and two possible binary measurements, $y\in\{0,1\}$ and $b\in\{0,1\}$. The score is given by 
\begin{equation}\label{Aeq}
\mathcal{A}_2 = \frac{1}{8}\sum_{x_0,x_1,y} P(b=x_y \lvert x_0,x_1,y) \,. 
\end{equation} 
This means that, upon receiving input $y$, the measurement device should return the output $b=x_y$, i.e.~the $y$-th bit of the input bit-string $x$ received by the preparation device. Hence the name of a $2\rightarrow 1$ RAC \cite{ANTV99,N99,TH15}. Note that all inputs are assumed to be chosen uniformly at random. Indeed, this task is nontrivial only when $d<4$; here we will consider the case $d=2$, i.e.~qubits. In this case, one finds the tight bounds $C_2 = 3/4$ and $Q_2 = (1+1/\sqrt{2})/2 \approx 0.85$ \cite{ANTV99}. The classical bound $C_2 $ can be obtained by simply always sending the bit $x_0$. The quantum bound $Q_2$ is obtained via the following ``ideal'' strategy. The four qubit preparations correspond to the pure states 
\begin{align}\nonumber
&\rho_{00}^{\text{ideal}}=\frac{\openone+\sigma_x}{2} &  \rho_{01}^{\text{ideal}}=\frac{\openone+\sigma_z}{2}\\
&\rho_{11}^{\text{ideal}}=\frac{\openone-\sigma_x}{2} &  \rho_{10}^{\text{ideal}}=\frac{\openone-\sigma_z}{2}.
\end{align}	
These are simply the eigenstates of the Pauli observables $\sigma_x$ and $\sigma_z$. Next, the measurements are projective and given by two anti-commuting Pauli observables 
\begin{equation}
M_y^{\text{ideal}} = (M_{y}^0)^{\text{ideal}} - (M_{y}^1)^{\text{ideal}}  = \frac{\sigma_x  + (-1)^y   \sigma_z}{\sqrt{2}}.
\end{equation}

These qubit preparations and measurements represent the ideal situation, where the maximal value $\mathcal{A}_2 = Q_2$ is achieved. In the following we will determine what restrictions apply to the possible preparations and measurements, given that a particular value of $\mathcal{A}_2$ is observed. In particular, when the maximal value $\mathcal{A}_2 = Q_2$ is attained, both the states and the measurements must be the ideal ones as given above (up to a unitary).

\textit{Self-testing preparations.---} Here we find restrictions on the set of prepared states given an observed value of $\mathcal{A}_2 $. For convenience, we write the qubit preparations as $\rho_{x_0x_1}=\left(\openone+\vec{m}_{x_0x_1}\cdot \vec{\sigma}\right)/2$, where $\vec{m}_{x_0x_1}$ denotes the Bloch vector (satisfying $\lvert \vec{m}_{x_0x_1}\rvert \leq 1$) and $\vec{\sigma} = (\sigma_x,\sigma_y, \sigma_z)$ denotes the vector of Pauli matrices. 

The first step consists in re-expressing   
\begin{equation}\label{rac2}
\mathcal{A}_2=\frac{1}{2}+\frac{1}{8}\sum_{y}\Tr\left(M_y^0V_y\right)
\leq \frac{1}{2}+\frac{1}{8}\sum_y \sqrt{\Tr\left(M_y^0V_y^2\right)\Tr\left(M_y^0\right)}
\end{equation}
where $V_{y}=\sum_{x_0,x_1}(-1)^{x_y}\rho_{x_0x_1}$. In the second step we used that for a positive semi-definite $O$ and a Hermitian operator $R$, it holds that $\lvert\Tr\left(OR\right)\rvert^2\leq \Tr\left(OR^2\right)\Tr\left(O\right)$ \cite{Kan17}. Without loss of generality, we can restrict to extremal qubit measurements, which are here projective rank-one operators. Consequently, we have that $\Tr\left(M_y^0\right)=1$. Next, we obtain $V_y^2 = \frac{1}{2}\left(\beta + (-1)^y \alpha\right)\openone$, where
\begin{align}
& \beta=\frac{1}{2}\sum_{x_0,x_1}\lvert \vec{m}_{x_0x_1}\rvert^2-\vec{m}_{00}\cdot \vec{m}_{11}-\vec{m}_{01}\cdot \vec{m}_{10} \\ &\alpha=\left(\vec{m}_{00}-\vec{m}_{11}\right)\cdot \left(\vec{m}_{01}-\vec{m}_{10}\right).
\end{align}
Finally, we find that Eq.~\eqref{rac2} reduces to
\begin{equation}\label{rac2a}
\mathcal{A}_2 \leq \frac{1}{2}+\frac{1}{8\sqrt{2}}\left[\sqrt{\beta+\alpha}+\sqrt{\beta-\alpha}\right].
\end{equation}

This provides a tight self-test of the prepared states (in terms of their Bloch vectors), for any given value of $\mathcal{A}_2$. Let us start with the case $\mathcal{A}_2 = Q_2$. Since $\sqrt{\beta+\alpha}+\sqrt{\beta-\alpha}=\sqrt{ 2\beta+2\sqrt{\beta^2-\alpha^2}}$, we see that Eq.~\eqref{rac2a} is maximal iff $\alpha=0$ and $\beta$ is maximal. This turns out to be achievable. In order to maximise $\beta$, we need (i) $\forall x_0x_1: \lvert \vec{m}_{x_0x_1}\rvert =1$, i.e.~that all preparations are pure states, and (ii) that $\vec{m}_{00}\cdot \vec{m}_{11}=\vec{m}_{01}\cdot \vec{m}_{10}=-1$, i.e.~the states correspond to (pairwise) antipodal Bloch vectors. We define $\vec{r}_0=\vec{m}_{00}=-\vec{m}_{11}$ and $\vec{r}_1=\vec{m}_{01}=-\vec{m}_{10}$. Consequently, we find $\alpha=4\vec{r}_0\cdot\vec{r}_1$. Therefore, in order to have $\alpha=0$, we must choose $\vec{r}_0\cdot \vec{r}_1=0$. This implies that the right-hand-side of Eq.~\eqref{rac2a} is upper-bounded by $Q_2$. Therefore, we conclude that when observing maximal value $\mathcal{A}_2 = Q_2$, the set of four prepared states must be equivalent (up to a unitary rotation) to the set of four ideal states; we note that this was also shown in Ref. \cite{Woodhead} in the context of QKD.

More generally, for any value $\mathcal{A}_2$, one can find a set of preparations (and corresponding measurements) such that the inequality \eqref{rac2a} is saturated, see Supplementary Material (SM) part A. For the case of classical preparations (i.e.~diagonal in a given basis), the Bloch vectors can simply be replaced by numbers $m_{x_0x_1} \in [-1,1]$, and we get $\mathcal{A}_2 \leq C_2$.

\textit{Self-testing measurements.---} Let us now consider self-testing of measurements. Using that $M_y = M_y^0 - M_y^1$, we write
\begin{align}\nonumber
 \mathcal{A}_2& =\frac{1}{2}+\frac{1}{16}\sum_{x_0,x_1} \Tr \big( \rho_{x_0x_1} [ (-1)^{x_0}M_{0}+(-1)^{x_1}M_{1}] \big)\\\label{meastest}
&\leq \frac{1}{2}+\frac{1}{16}\sum_{x_0,x_1}  \lmax \left[(-1)^{x_0}M_{0}+(-1)^{x_1}M_{1}\right],
\end{align}
where $\lmax[X]$ is the largest eigenvalue of the (Hermitian) operator $X$. Since the upper bound corresponds to choosing the optimal preparations for a fixed pair of observables, it simply quantifies the optimal performance achievable using these observables. If $M_{0}$ and $M_{1}$ are qubit observables the upper bound can be evaluated exactly (see SM part A) to give
\begin{align}
\label{eq:observables-upper-bound}
\mathcal{A}_2 \leq \frac{1}{2} + \frac{1}{16} \bigg( \sqrt{ 2 \mu + 2 \nu - \eta_{+}^{2} } + \sqrt{ 2 \mu - 2 \nu - \eta_{-}^{2} } \bigg),
\end{align}
where $\mu = \Tr \big( M_{0}^{2} + M_{1}^{2} \big)$, $\nu = \Tr \{M_{0}, M_{1} \}$ and $\eta_{\pm} = \Tr (M_{0} \pm M_{1})$. The right-hand side reaches the optimal value $Q_{2}$ iff $\mu = 4, \eta_{\pm} = 0$ and $\nu = 0$, which implies anti-commuting projective observables (i.e. projective measurement operators). In other words, observing $\mathcal{A}_2 = Q_2$ implies that the measurements are unitarily equivalent to the ideal ones. Moreover, note that inequality \eqref{eq:observables-upper-bound} is tight; for any value of $\mathcal{A}_2$ one can find measurements (and corresponding states) such that inequality is saturated (see SM part A). It follows that any pair of projective, rank-one observables that is incompatible ($\lvert \nu \rvert < 4$) can lead to $\mathcal{A}_2>C_2$.

\emph{Robust self-testing of the preparations.---} 
We now discuss the problem of characterizing the fidelity between the realised preparations and the ideal ones. This will allow us to quantify the distance of the prepared states with respect to the ideal ones. Again, we want to develop self-testing methods which are based only on the value of $\mathcal{A}_2$.

More formally, given an arbitrary set of preparations, we define the average fidelity with the ideal preparations to be $S(\{\rho_{x_0x_1}\})=\max_{\Lambda}\sum_{x_0,x_1} F(\rho_{x_0x_1}^{\text{ideal}},\Lambda[\rho_{x_0x_1}])/4$, where $\Lambda$ is a quantum channel i.e.~a completely positive trace-preserving map. Here the fidelities $F(\rho,\sigma)=\Tr\left(\sqrt{\sqrt{\rho}\sigma\sqrt{\rho}}\right)$ simplify to $F(\rho^{\text{ideal}}_{x_0x_1},\Lambda[\rho_{x_0x_1}])=\Tr\left(\Lambda[\rho_{x_0x_1}]\rho^{\text{ideal}}_{x_0x_1}\right)$, as the $\rho^{\text{ideal}}_{x_0x_1}$ are pure states.  We derive lower bounds on the smallest possible value of $S$ given a value of $\mathcal{A}_2$,	 i.e.,
\begin{equation}\label{fid}
\mathcal{F}\left(\mathcal{A}_2\right)= \min_{ \{ \rho_{x_0x_1}\}\in R(\mathcal{A}_2)} S\left[\{\rho_{x_0x_1}\}\right]. 
\end{equation}
Note that this involves a minimisation over all sets of four preparations $R(\mathcal{A}_2)$ that are compatible with an observed value $\mathcal{A}_2$.

In order to lower bound $\mathcal{F}$, we use an approach inspired by Ref.~\cite{Kan16}. From Eq.~\eqref{meastest}, we have $\mathcal{A}_2= \frac{1}{2}+\sum_{x_0,x_1}\Tr\left(W_{x_0x_1}\rho_{x_0x_1}\right)$, where  $W_{x_0x_1}=\frac{1}{16}\sum_y(-1)^{x_y}M_y$. 
We define operators corresponding to some suitably chosen channel acting on the ideal preparations:
\begin{equation}\label{channel}
K_{x_0x_1}(M_0,M_1)= \Lambda^\dagger(M_0,M_1)[\rho^{\text{ideal}}_{x_0x_1}],
\end{equation}
where $\Lambda^\dagger$ is the channel dual to $\Lambda$. We aim to construct operator inequalities of the form 
\begin{align}\label{op}
K_{x_0x_1}(M_0,M_1)\geq s W_{x_0x_1}+t_{x_0x_1}(M_0,M_1)\openone,
\end{align}
for all inputs $(x_0,x_1)$, for any given measurements, where $s$ and $t_{x_0x_1}(M_0,M_1)$ are real coefficients. Finding such inequalities, as well as a suitable channel $\Lambda$ allows us to lower bound $S$ as follows: 

\begin{multline}\label{fidllow}
S \geq\frac{1}{4}\sum_{x_0,x_1}\Tr\left(K_{x_0x_1}\rho_{x_0x_1}\right)\geq \frac{s}{4}\sum_{x_0,x_1}\Tr\left(W_{x_0x_1}\rho_{x_0x_1}\right)\\
+\frac{1}{4}\sum_{x_0,x_1}t_{x_0x_1}=\frac{s}{4}(\mathcal{A}_2-1/2)+\frac{1}{4} \sum_{x_0,x_1}t_{x_0x_1},
\end{multline}
Applying a minimisation over $M_0$ and $M_1$ to the right-hand-side, the above inequality becomes valid for all preparations. Consequently,  
\begin{equation}\label{fidlow}
\mathcal{F}(\mathcal{A}_2)\geq \frac{s}{4}\left(\mathcal{A}_2-1/2\right)+t\equiv L\left(\mathcal{A}_2\right),
\end{equation}
where $t \equiv 1/4  \min_{M_0,M_1}\sum_{x_0,x_1} t_{x_0x_1}(M_0,M_1)$. In SM part B, we construct explicitly the channel and derive an operator inequality leading to a lower bound, given by $ s=4\left(1+\sqrt{2}\right)$ and $t=\left(2-\sqrt{2}\right)/4$. 

This provides a robust self-testing for the preparations. A maximal value $\mathcal{A}_2 = Q_2$ implies $\mathcal{F}=1$, i.e.~the preparations must be the ideal ones (up to a unitary). For $\mathcal{A}_2 = C_2$, i.e.~a maximal value given a set of classical states, we get that $\mathcal{F} \geq 3/4$. This bound can be attained via the set of pure states $\rho_{x_0x_1} = (\openone  + (-1)^{x_0x_1} \sigma_z)/2$ (diagonal in the same basis, hence classical), combined with the measurements $M_{0}= M_1 = \sigma_z$. Therefore, we see that our bound $\mathcal{F}\left(\mathcal{A}_2\right) \geq L(\mathcal{A}_2)$ is optimal, as far as linear inequalities are concerned (see Fig.~1). It is then interesting to consider the intermediate region $C_2 <\mathcal{A}_2 < Q_2$. First, focusing on strategies involving a single set of states and measurements, we observe numerically that the linear bound $\mathcal{F}\left(\mathcal{A}_2\right) \geq L(\mathcal{A}_2)$ cannot be saturated anymore, and conjecture the form of optimal states and measurements; see red curve in Fig.~1 and SM part C for details. Second, allowing for shared randomness between the preparation and measurement device (such that convex combinations of qubit strategies are now possible), the linear bound becomes tight, a direct consequence of the linearity of $\mathcal{F}$ and $\mathcal{A}_2$ in terms of the states and measurements.

\begin{figure}[t!]
	\centering
	\includegraphics[width=0.9\columnwidth]{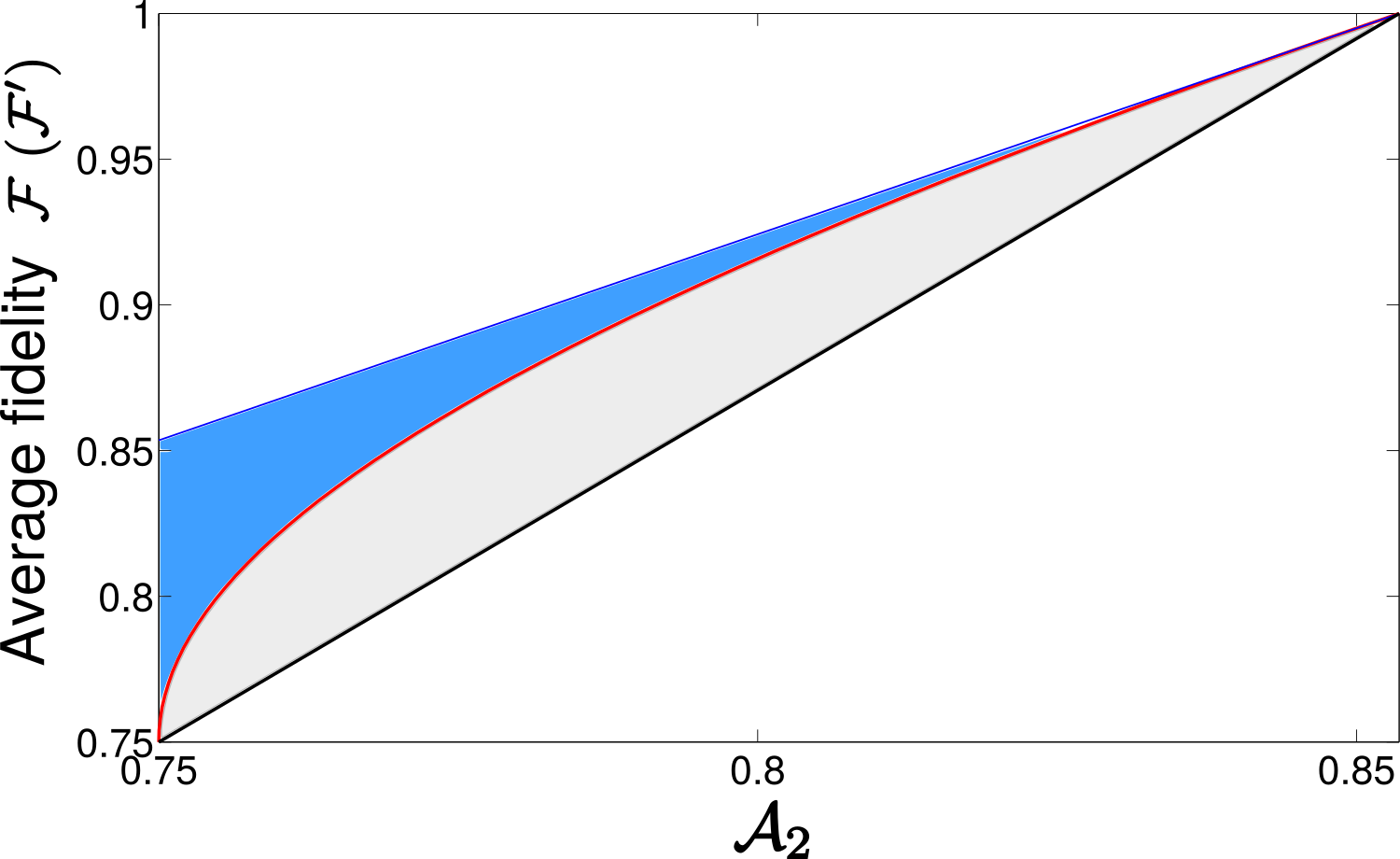}
	\caption{Average fidelity $\mathcal{F}$ $(\mathcal{F}')$ for prepared states (measurements), as a function of the observed value of $\mathcal{A}_2$. The black line is our analytical lower bound of Eq.~\eqref{fidlow}. The blue region is accessible via single qubit strategies without shared randomness, as confirmed by strong numerical evidence (see SM). When allowing for shared randomness between the devices, the accessible region (obtained by taking the convex hull of the blue region) now also includes the grey area, and our analytic lower bound is tight in general.}\label{fig1}
\end{figure}

\emph{Robust self-testing of the measurements.---} 
Similarly, we can quantify the average fidelity of the measurements with respect to the ideal ones: $S'\left(\{M_y^b\}\right)=\max_{\Lambda}\sum_{y,b} F\left((M_y^b)^{\text{ideal}},\Lambda[M_y^b]\right)/4$, where $\Lambda$ must be a unital channel (i.e. mapping the identity to itself), in order to ensure that measurements are mapped to measurements. In analogy with the case of preparations, our goal is to lower bound the following quantity:
\begin{equation}
\mathcal{F}'(\mathcal{A}_2)= \min_{ \{M_y^b\} \in R'(\mathcal{A}_2)}  S'\left(\{M_y^b\}\right),
\end{equation}
where $R'(\mathcal{A}_2)$ represents all sets of measurements compatible with a certain value of $\mathcal{A}_2$.

We first rewrite $\mathcal{A}_2=\sum_{y,b} \Tr (M_y^b Z_{yb} )$, where $Z_{yb}= \frac{1}{8}\sum_{x_0,x_1} \rho_{x_0x_1}\delta_{b,x_y}$. 
Next, we construct operator inequalities 
\begin{equation}
K_{yb}(\{\rho_{x_0x_1}\})\geq sZ_{yb}+t_y(\{\rho_{x_0x_1}\})\openone \,,
\end{equation}
given the unital channel $K_{yb}=\Lambda^\dagger[(M_y^b)^{\text{ideal}}]$. Similarly to the case of preparations, strong operator inequalities can be derived by choosing carefully the channel; all details are given in SM part D. Finally, this leads to a lower bound on the average fidelity
\begin{align}\label{Fprime}
\mathcal{F}'(\mathcal{A}_2) & \geq \min_{\{\rho_{x_0x_1}\}}\frac{1}{4}\sum_{y,b}\Tr\left(K_{yb}M_y^b\right)  \geq L(\mathcal{A}_2).
\end{align}
That is, we find that $\mathcal{F}'$ can be lower-bounded by a linear expression in terms of $\mathcal{A}_2$, which turns out to be the same as for the case of preparations. 

This provides a robust self-test for the measurements. Observing $\mathcal{A}_{2}=Q_2$ implies that $\mathcal{F}'=1$, hence the measurements are equivalent to the ideal ones (up to a unitary). For $\mathcal{A}_{2}=C_2$, we have that $\mathcal{F} \geq 3/4$. This lower bound can be attained by choosing $M_0 = \sigma_z$ and $M_1 = \openone$, with the states $\rho_{00}=\rho_{01} = (\openone+ \sigma_z)/2$ and $\rho_{10}=\rho_{11} = (\openone- \sigma_z)/2$. For $C_2<\mathcal{A}_{2}<Q_2$, we find numerically that the inequality \eqref{Fprime} cannot be saturated using a single set of measurements and states (see Fig.~1). Details, in particular a conjecture for the form of the optimal measurements, are given in SM part C. Similarly as for the case of states, when allowing for convex combinations of qubit strategies, our linear bound is tight.

\textit{Generalisations.---} The above results can be generalised in several directions. Firstly, a generalisation of the $2 \rightarrow 1$ RAC enables self-testing of any pair of incompatible Pauli observables (see SM part E). Secondly, we derive compatibility relations for the $N \rightarrow 1$ RAC with qubits (see SM part F). Thirdly, we self-test qutrit preparations and projective measurements in the $2 \rightarrow 1$ RAC (see SM part G).

Finally, we present a numerical method for robust self-testing of preparations applicable in scenarios beyond RACs. The method is based on semi-definite programing and combines (i) the swap-method \cite{swaptrick} used for self-testing in Bell scenarios with (ii) the hierarchy of finite-dimensional quantum correlations \cite{NV, NV2, Symmetry}. The idea is to first construct a swap operator, based on the measurement operators, which maps the state of the preparation onto an ancilla. The average fidelity between the ancilla and the ideal states can then be expressed in terms of strings of products of measurements operators and the extracted states. The last step is to miminize this average fidelity over all quantum realisations that are compatible with a given witness value, using the hierarchy of Refs \cite{NV, NV2, Symmetry}. Although typically returning sub-optimal bounds on $\mathcal{F}$, this method is widely applicable. In SM part H, we describe in detail the methodology and apply to two examples, including the $2 \rightarrow 1$ RAC.


\textit{Outlook.---} We presented methods for self-testing quantum states and measurements in the prepare-and-measure scenario. These techniques demonstrate strong robustness to noise, and should therefore be directly amenable to experiments, providing useful certification techniques in a semi-device-independent setting. Moreover, these ideas should find applications in quantum communications. Our methods apply to the states and measurements used in QKD (e.g. in BB84), as well as in semi-device-independent QKD and randomness generation protocols \cite{PB11,LYW11,WP15,Lunghi,MT16}.

It would be interesting to develop robust self-testing techniques for more general scenarios, e.g. for higher dimensional quantum systems. Another direction would be to consider scenarios beyond prepare-and-measure, for instance adding between the preparation and measurement devices a transformation device \cite{TS05, bowles15}, and self-test the latter.

Finally, while we have focused here on self-testing based on an assumption on the dimension, one could develop methods based on different assumptions, such as a bound on the mean energy \cite{thomas}, the overlap \cite{BME17}, or the entropy \cite{chaves}.

\textit{Acknowledgements.---}
This work was supported by the Swiss national science foundation (Starting grant DIAQ, QSIT, and Early Postdoc Mobility fellowship P2GEP2\_162060), the European Union's Horizon 2020 research
and innovation programme under the Marie Sk{\l}odowska-Curie Action
ROSETTA (grant no.~749316), the European Research Council (grant
no.~337603), the Danish Council for Independent Research (Sapere Aude), VILLUM FONDEN via the QMATH Centre of Excellence (grant no.~10059), and the National Research, Development and Innovation Office NKFIH (Grants No. K111734, and No. KH125096).

\appendix
\onecolumngrid

\section{Self-testing relations for preparations and measurements}\label{A}

In this section we provide a simple example of preparations that saturate the compatibility bound for $\mathcal{A}_2$ given in the main text. Moreover, we derive the upper bound for compatibility of measurements given in the main text.

First, we consider the case of preparations. 
Consider preparations such that $\rho_{00}$ and $\rho_{11}$, and $\rho_{01}$ and $\rho_{10}$ correspond to antipodal Bloch vectors with a relative angle $\theta$, the maximal quantum value of $\mathcal{A}_2$, is obtained from
\begin{equation}\label{A2opt}
\mathcal{A}_2=\frac{1}{2}+\frac{1}{8}\sum_{y}\lmax\left[V_y\right],
\end{equation}
where $V_y=\sum_{x_0,x_1}(-1)^{x_y}\rho_{x_0x_1}$. We represent the preparations on the Bloch sphere as $\rho_{x_0x_1}=1/2\left(\openone+\vec{m}_{x_0x_1}\cdot\vec{\sigma}\right)$, where $\vec{m}_{00}=\left[\cos(\theta/2),0,\sin(\theta/2)\right]$ and $\vec{m}_{01}=\left[\cos(\theta/2),0,-\sin(\theta/2)\right]$, with $\vec{m}_{11}=-\vec{m}_{00}$ and $\vec{m}_{10}=-\vec{m}_{01}$. This gives $V_0=2\cos(\theta/2)\sigma_x$ and $V_1=2\sin(\theta/2)\sigma_z$. The respective largest eigenvalues are $\lmax\left[V_0\right] = 2\cos(\theta/2)$ and $\lmax\left[V_1\right] =2\sin(\theta/2)$, leading to 
\begin{equation}
\mathcal{A}_2= \frac{1}{2}+ \frac{1}{4\sqrt{2}} \left[\sqrt{1+\cos\theta}+\sqrt{1-\cos\theta}\right].
\end{equation} 
It is straightforward to see that this achieves the upper bound in the main text; indeed the above choice of preparations leads to $\beta=4$ and $\alpha=4\cos\theta$.

%
In order to derive the upper bound on $\mathcal{A}_2$ for compatibility of measurements in the main text we evaluate
\begin{equation}
\sum_{x_{0} ,x_{1}} \lmax \big[ (-1)^{x_0} M_{0}+(-1)^{x_1} M_{1} \big]
\end{equation}
for arbitrary qubit observables $M_{0}, M_{1}$. We take advantage of the fact that
\begin{equation}
\lmax [ T ] + \lmax [ -T ] = \lmax [ T ] - \lambda_{\textnormal{min}} [ T ],
\end{equation}
which for a $2 \times 2$ matrix can be evaluated analytically. More specifically, if $T$ is a $2 \times 2$ Hermitian matrix with eigenvalues $\lambda_{0} \geq \lambda_{1}$, let
\begin{align*}
\chi &:= \Tr T = \lambda_{0} + \lambda_{1},\\
\zeta &:= \Tr T^{2} = \lambda_{0}^{2} + \lambda_{1}^{2}
\end{align*}
and then
\begin{equation}
\label{eq:eigenvalue-difference}
\lambda_{0} - \lambda_{1} = \sqrt{ 2 \zeta - \chi^{2} }.
\end{equation}
Evaluating this expression for $T = M_{0} \pm M_{1}$ gives the desired upper bound given.

\section{Operator inequalities for robust self-testing of preparations}\label{B}

In this section we provide a detailed derivation of the lower bound on the average fidelity $\mathcal{F}(\mathcal{A}_2)$.

For a real constant $s > 0$, to be chosen later, consider for each pair $(x_0, x_1)$ the operator $K_{x_0 x_1} - s W_{x_0 x_1}$, where $W_{x_0x_1}=\frac{1}{16}\sum_y(-1)^{x_y}M_y$ and $K_{x_0x_1}= \Lambda^\dagger[\rho^{\text{ideal}}_{x_0x_1}]$, for some channel $\Lambda$.
Suppose now that $t_{x_0 x_1} \in \mathbb{R}$ is a lower bound on its eigenvalues, or, equivalently, that the operator inequality
\begin{equation}
\label{op3}
K_{x_0 x_1} \geq s~W_{x_0x_1} + t_{x_0x_1}~\openone
\end{equation}
holds. Then, computing the trace of this inequality with $\rho_{x_0 x_1}$ and averaging over inputs leads to
\begin{equation}
S \ge \frac{1}{4} \sum_{x_0, x_1} F(\rho^\text{ideal}_{x_0 x_1}, \Lambda[\rho_{x_0 x_1}]) \ge \frac{s}{4}\left (\mathcal{A}_2 - \frac{1}{2} \right ) + t, \qquad t \equiv \frac{1}{4} \sum_{x_0, x_1} t_{x_0 x_1} \;,
\end{equation}
where the first inequality holds because $S$ is defined as maximisation over all possible channels, and the $\Lambda$ used there is one possible choice.
In turn, if~\eqref{op3} holds {\em as an operator inequality}, it is valid for any set of preparations $\{ \rho_{x_0 x_1} \}$, and thus $\mathcal{F}(\mathcal{A}_2) \ge \frac{s}{4} \left( \mathcal{A}_2 - \frac{1}{2} \right) + t$. Note that~\eqref{op3} has a dependence on $M_0$, $M_1$ through $W_{x_0 x_1}$. If~\eqref{op3} holds for a particular choice of measurement operators $M_0$, $M_1$, then the bound on $\mathcal{F}(\mathcal{A}_2)$ holds for all preparations, for that particular choice of $M_0$, $M_1$. However, if~\eqref{op3} holds for {\em all} possible $M_0$, $M_1$, then the bound on $\mathcal{F}(\mathcal{A}_2)$ is valid for all quantum setups and is thus a robust self-testing inequality. To derive the appropriate constants $s$ and $t_{x_0 x_1}$, we first allow $t_{x_0 x_1}$ and $\Lambda$ to have a dependence on $M_0$ and $M_1$. We then minimise over $M_0$ and $M_1$ the constants $t_{x_0 x_1}$, for a suitable choice of $s$, such that, at the end, \eqref{op3} holds regardless of the choice of measurement operators.

We choose a dephasing channel of the form
\begin{equation}
\Lambda_\theta(\rho)=\frac{1+c(\theta)}{2}\rho+\frac{1-c(\theta)}{2}\Gamma(\theta)\rho\Gamma(\theta),
\end{equation}
where for $0\leq \theta\leq \pi/4$ we use $\Gamma=\sigma_x$, while for $\pi/4< \theta\leq \pi/2$ we use $\Gamma=\sigma_z$. The function $c(\theta) \in [-1, 1]$ will be specified later.

In the interval $0\leq \theta\leq \pi/4$, the action of the channel leads to
\begin{align}
& K_{00}=\frac{\openone+\sigma_x}{2}  &  K_{01}=\frac{\openone+c(\theta)\sigma_z}{2}  &&  K_{10}=\frac{\openone-c(\theta)\sigma_z}{2}  &&   K_{11}=\frac{\openone-\sigma_x}{2},
\end{align}
whereas in the interval $\pi/4< \theta\leq \pi/2$, we have 
\begin{align}
& K_{00}=\frac{\openone+c(\theta)\sigma_x}{2} &  K_{01}=\frac{\openone+\sigma_z}{2} && K_{10}=\frac{\openone-\sigma_z}{2} && K_{11}=\frac{\openone-c(\theta)\sigma_x}{2}.
\end{align}

As discussed in the main text, for any given set of preparations, the optimal measurements are projective and rank-one. Furthermore, any two such measurements can be represented on an equator of the Bloch sphere. Due to the freedom of setting the reference frame, we can without loss of generality represent the two measurements in the $xz$-plane, i.e.
\begin{align}
& M_0=\cos\theta \hspace{1mm}\sigma_x+\sin \theta\hspace{1mm}\sigma_z &  M_1=\cos \theta \hspace{1mm}\sigma_x-\sin \theta\hspace{1mm}\sigma_z \nonumber.
\end{align}
We can therefore write $W_{x_0x_1}$ as
\begin{align}
& W_{00}=\frac{1}{8}\cos\theta \sigma_x & W_{01}=\frac{1}{8}\sin\theta \sigma_z && W_{10}=-\frac{1}{8}\sin\theta \sigma_z && W_{11}=-\frac{1}{8}\cos\theta \sigma_x.
\end{align}
We can reduce the number of operator inequalities \eqref{op3} by exploiting the apparent symmetries in the expressions for $W_{x_0x_1}$ and $K_{x_0x_1}$: we restrict ourselves so that $t_o\equiv t_{01}=t_{10}$ and $t_e\equiv t_{00}=t_{11}$. Thus, we have to consider two operator inequalities in each interval $\theta\in [0,\pi/4]$ and $\theta\in (\pi/4,\pi/2]$. In the first interval, the two operator inequalities are
\begin{align}
& \frac{1+\sigma_x}{2}-\frac{s}{8}\cos\theta\sigma_x-t_e\openone\geq 0 &
& \frac{1+c(\theta)\sigma_z}{2}-\frac{s}{8}\sin\theta\sigma_z-t_o\openone\geq 0.
\end{align}
In the second interval, the two operator inequalities are
\begin{align}
& \frac{1+c(\theta)\sigma_x}{2}-\frac{s}{8}\cos\theta\sigma_x-t_e\openone\geq 0 &
& \frac{1+\sigma_z}{2}-\frac{s}{8}\sin\theta\sigma_z-t_o\openone\geq 0.
\end{align}
We now focus on the former interval. Solving the two inequalities for $t_o$ and $t_e$ we obtain
\begin{align}
&t_e\leq 1-\frac{s}{8}\cos\theta, &t_o&\leq \frac{1}{8}\left(4+4c(\theta)-s\sin\theta \right),\\
&t_e\leq \frac{s}{8}\cos\theta, &t_o&\leq \frac{1}{8}\left(4-4c(\theta)+s \sin\theta \right).
\end{align}
Any choice of $t_o$ and $t_e$ satisfying these constraints gives rise to valid operator inequalities. In order to obtain the strongest bound, we choose the largest values of $t_o$ and $t_e$ consistent with their respective constraints, i.e.,
\begin{align}\label{eq1}
& t_e=\min\left\{1-\frac{s}{8}\cos\theta ,\frac{s}{8}\cos\theta\right\} & t_o=\min \left\{ \frac{1}{8}\left(4+4c(\theta)-s\sin\theta \right),\frac{1}{8}\left(4-4c(\theta)+s \sin\theta \right)\right\}.
\end{align}
A similar procedure for the interval $\theta\in (\pi/4,\pi/2]$ leads to 
\begin{align}\label{eq2}
& t_e=\min \left\{ \frac{1}{8}\left(4+4c(\theta)-s\cos\theta \right),\frac{1}{8}\left(4-4c(\theta)+s \cos\theta \right)\right\} & t_o=\min\left\{1-\frac{s}{8}\sin\theta ,\frac{s}{8}\sin\theta\right\}.
\end{align}
It is worth pointing out that the two intervals only differ by exchanging $t_{e} \leftrightarrow t_{o}$ and $\sin \theta \leftrightarrow \cos \theta$. Hence, for any given $\theta$, we have constructed operator inequalities of the form \eqref{op3}. 

As shown in the main text, we obtain our lower bound on the average fidelity from
\begin{equation}\label{fidlow2}
\mathcal{F}\left(\mathcal{A}_2\right)\geq \frac{s}{4}(\mathcal{A}_2-1/2)+\min_{M_0,M_1}t(M_0,M_1)\equiv L(\mathcal{A}_2),
\end{equation}
where $t(M_0,M_1)=\left(t_e+t_o\right)/2$. To compute this quantity we fix the value of $s$ to be
\begin{equation}
s=4\big(1+\sqrt{2}\big)
\end{equation}
and choose the dephasing function as $c(\theta)=\min\{1,\frac{s}{4}\sin\theta\}$ whenever $\theta\in [0,\pi/4]$, and $c(\theta)=\min\{1,\frac{s}{4}\cos\theta\}$ whenever $\theta\in (\pi/4,\pi/2]$. It is easy to see that $c(\theta) \in [0, 1]$, which ensures that $\Lambda_{\theta}$ is a valid quantum channel, and that $c(\theta)$ is continuous at $\theta=\pi/4$. A simple calculation shows that in this case
\begin{equation}
t=\frac{2-\sqrt{2}}{4},
\end{equation}
which gives the lower bound
\begin{equation}
\mathcal{F}(\mathcal{A}_2)\geq \left(1+\sqrt{2}\right)\mathcal{A}_2-\frac{3}{2\sqrt{2}} \equiv L(\mathcal{A}_2).
\end{equation}
One can check that choosing distinct values of $s$ will not lead to improved lower bounds.

\section{Tightness of fidelity bounds}\label{Cbis}

In the main text, we have derived fidelity bounds for both the preparations and the measurements, based on operator inequalities. Specifically, we obtain a lower bound on the average fidelity $\mathcal{F}$ of the prepared states (with respect to the ideal ones) given by the linear expression
\begin{equation}\label{F_A2}
\mathcal{F}(\mathcal{A}_2)\geq \left(1+\sqrt{2}\right)\mathcal{A}_2-\frac{3}{2\sqrt{2}} \equiv L(\mathcal{A}_2).
\end{equation}
For measurements, a similar bound is obtained on the average fidelity $\mathcal{F}'$ with respect to the ideal ones. In the present appendix, we discuss the tightness of these bounds. 

We start with our bound on the fidelity of the states. As discussed in the main text, obtaining $\mathcal{A}_2 = Q_2$ implies $\mathcal{F}=1$, i.e. the states are the ideal ones (up to a unitary). Let us refer to the optimal strategy (with the ideal states) as strategy $S_1$. Then, for $\mathcal{A}_2 = C_2$, our bound gives $\mathcal{F} \geq 3/4$. This bound is tight and can be obtained via the set of pure states $\rho_{x_0x_1} = (\openone  + (-1)^{x_0x_1} \sigma_z)/2$ (diagonal in the same basis, hence classical), combined with the measurements $M_{0}= M_1 = \sigma_z$. Let us refer to this strategy as $S_2$. 

The above shows that our bound \eqref{F_A2} is tight as far as linear inequalities are concerned. More generally, the bound is in fact tight in general, when shared randomness between the preparation and measurement devices is taken into account. In this case, taking a convex combination between strategies $S_1$ and $S_2$ allows us to get any point on the line (i.e. pair of values $\mathcal{F}$ and $\mathcal{A}_2$) between $S_1$ and $S_2$.

\begin{figure}[t!]
	\centering
	\includegraphics[width=0.6\columnwidth]{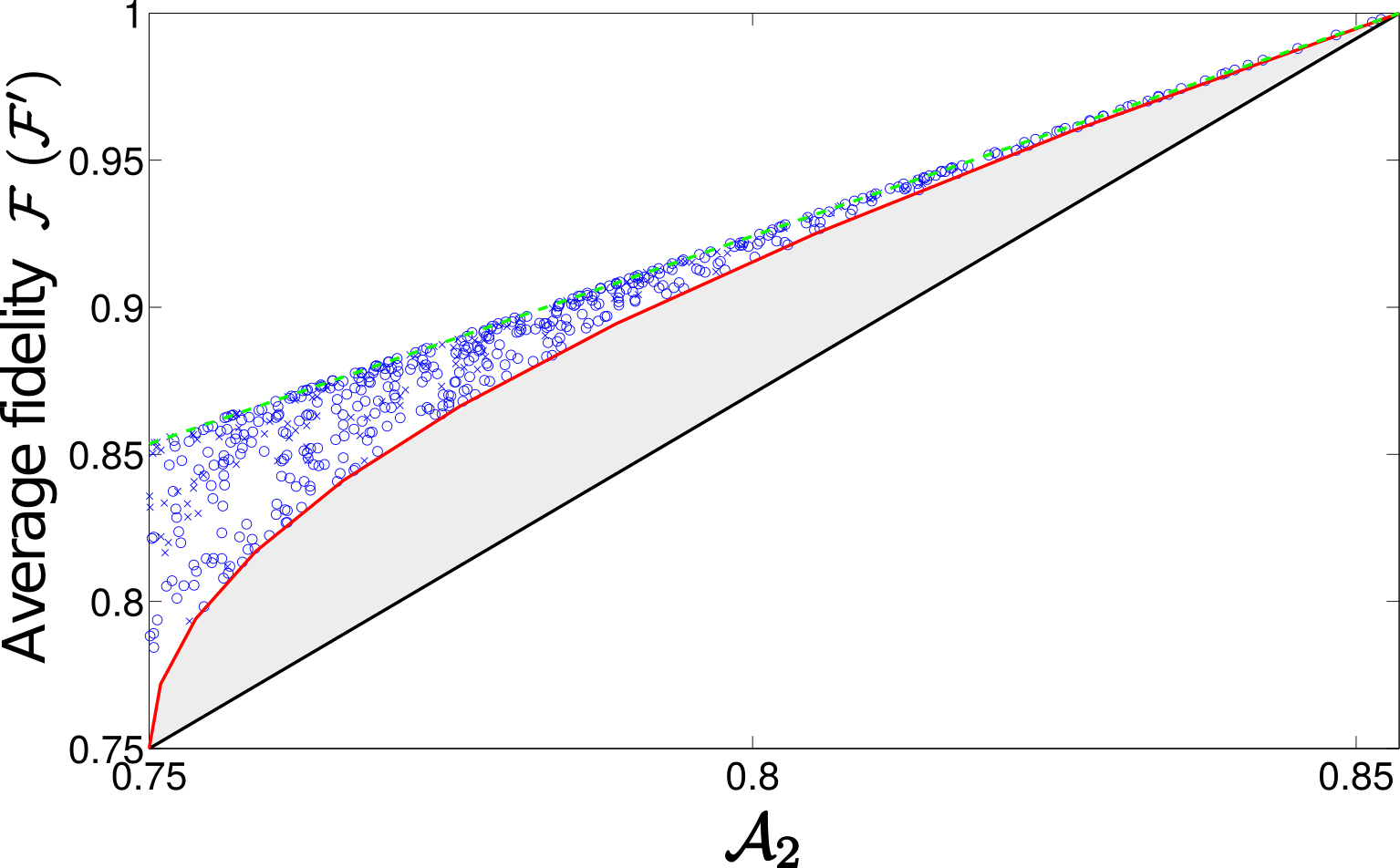}
	\caption{The black line is the analytic lower bound on the average fidelity $\mathcal{F}$ $(\mathcal{F}')$ for prepared states (measurements), as a function of the observed value of $\mathcal{A}_2$. To characterise the region accessible via pure qubit strategies (i.e. without shared randomness), we perform numerics generating randomly sets of qubit preparations (blue circles and crosses); here we show the numerical results for the case of states, but similar results are obtained for the case of measurements. In the region $C_2 <\mathcal{A}_2 < Q_2$, we conjecture that the class of strategies given in the text (corresponding to the red curve) are optimal; both for $\mathcal{F}$ and $\mathcal{F}'$. Finally, the green dashed line is our conjectured upper bound on the average fidelity.}\label{fig2}
\end{figure}

It is also interesting to understand what happens when shared randomness between the devices is not taken into account. In this case, the end points ($\mathcal{A}_2 = Q_2, \mathcal{F}=1$) and ($\mathcal{A}_2 = C_2, \mathcal{F}=3/4$) can still be obtained. To understand what happens in the intermediate region $C_2 <\mathcal{A}_2 < Q_2$, we first performed a numerical analysis. Specifically, we chose randomly four qubit states, and compute (i) the maximal value of $\mathcal{A}_2$ (optimizing over the measurements), and (ii) the average fidelity $\mathcal{F}$ (where the optimization over channels is restricted here to unitaries). The resulting points are shown on Fig.~2 (blue circles). This indicates that for $C_2 <\mathcal{A}_2 < Q_2$, the bound \eqref{F_A2} cannot be saturated anymore. Moreover, we conjecture that an optimal class of strategies is given by the pure states
\begin{equation}
\ket{\psi_{00}} = \ket{0} \quad \ket{\psi_{11}} = \ket{1} \quad \ket{\psi_{01}} =  \cos{\theta} \ket{0} + \sin{\theta} \ket{1}
\quad \ket{\psi_{10}} =  \cos{\theta} \ket{0} - \sin{\theta} \ket{1}  \,
\end{equation}
and the measurements $M_y = \cos(\varphi) \sigma_z  + (-1)^y \sin(\varphi) \sigma_x  $. Straightforward calculations show that, taking $\tan{\varphi}= \sin{2\theta}$, leads to 
\begin{equation}
\mathcal{A}_2 = \frac{1}{2} + \frac{1}{4} \sqrt{1 + \tan^2(\varphi)}
\quad   \textrm{and} \quad \mathcal{F} = \frac{1}{4} (3 + \tan{\varphi} )  \,.
\end{equation}
This gives a parametric curve, as a function of $\varphi \in [0 , \pi/4]$, given by the red curve in Fig.~2. This  curve is in excellent agreement with the numerical results obtained before. Note that this class of strategies interpolates between the strategies $S_1$ (setting $\varphi =0 $) and $S_2$ (setting $\varphi = \pi/4$).

Next we discuss the bound on the average fidelity of measurements. As discussed in the main text, the linear bound $\mathcal{F}'(\mathcal{A}_2)\geq  L(\mathcal{A}_2)$ is optimal as far as linear inequalities are concerned. Moreover, when allowing for shared randomness the bound is tight in general for $C_2 \leq\mathcal{A}_2 \leq Q_2$. This is obtained by considering convex combinations of strategy $S_1'$ (defined as the optimal strategy $S_1$, up to a rotation of $\pi/8$ around the $y$ axis; see below), and the following strategy (referred to as $S_3$): take $M_0 = \sigma_z$ and $M_1 = \openone$, with the states $\rho_{00}=\rho_{01} = (\openone+ \sigma_z)/2$ and $\rho_{10}=\rho_{11} = (\openone- \sigma_z)/2$. 

Similarly to the case of states, we now consider the situation where shared randomness between the devices is not allowed. Performing a numerical analysis similar to the one described above (except that measurements are no generated randomly), we observe that the accessible region (in terms of $\mathcal{F}'$ vs $\mathcal{A}_2$) appears to be exactly the same as for the case of states (i.e. the blue region in Fig.~2). We conjecture that the lower bound is given by the following class of optimal strategies: take the measurements:
\begin{equation}
M_0 = \sigma_z    \quad \textrm{and} \quad   M_1 = \eta \sigma_x + (1-\eta) \openone
\end{equation}
with the states $\ket{\psi_{00}} =  \cos{\theta} \ket{0} + \sin{\theta} \ket{1}$, $\ket{\psi_{01}} =  \cos{\theta} \ket{0} - \sin{\theta} \ket{1}$, $\ket{\psi_{10}} =  \cos{\theta} \ket{1} + \sin{\theta} \ket{0}$, and $\ket{\psi_{11}} =  \cos{\theta} \ket{1} - \sin{\theta} \ket{0}$. Setting $\eta = \tan{2 \theta}$, we get
\begin{equation}
\mathcal{A}_2 = \frac{\cos^2(\theta)}{2} + \frac{1}{4} + \frac{\sin^2(2\theta)}{\cos(2\theta)}
\quad   \textrm{and} \quad \mathcal{F} = \frac{1}{4} (3 + \tan(2\theta) )  \,.
\end{equation}
This gives a parametric curve, as a function of $\theta \in [0 , \pi/8]$, given by the red curve in Fig.~2. This  curve is in excellent agreement with the numerical results obtained before. Also, this curve turns out to be exactly the same as the curve we obtained above for the case of states. Note that this class of strategies interpolates between the strategies $S_1'$ (setting $\theta =\pi/8 $) and $S_3$ (setting $\theta = 0$).

Finally, note that the numerics also suggests that there is a linear upper bound on the average fidelities $\mathcal{F}$ ($\mathcal{F}'$) as a function of $\mathcal{A}_2$ (see Fig.~2); specifically $\mathcal{F} \leq \frac{1-Q_2}{Q_2 - 3/4} \mathcal{A}_2 + \frac{Q_2^2 -3/4}{Q_2 - 3/4}$ and similarly for $\mathcal{F}'$. It would be interesting to provide a proof of these upper bounds.

\section{Operator inequalities for robust self-testing of measurements}\label{C}

In this section, we account for the detailed derivation of the lower bound on the average fidelity of the measurements $\mathcal{F}'(\mathcal{A}_2)$. The approach bears significant resemblance to the case of robustly self-testing preparations, as outlined in Appendix \ref{B}. 

We aim to derive operator inequalities of the form
\begin{equation}\label{opm}
K_{yb}(\{\rho_{x_0x_1}\})\geq sZ_{yb}+t_y(\{\rho_{x_0x_1}\})\openone,
\end{equation}
where  $Z_{yb}= \frac{1}{8}\sum_{x_0,x_1} \rho_{x_0x_1}\delta_{b,x_y}$ and $K_{yb}(\{\rho_{x_0x_1}\})=\Lambda^\dagger[(M_y^b)^{\text{ideal}}]$. For the sake of simplicity, we first apply a unitary channel to $(M_y^b)^{\text{ideal}}$ align these operators with the eigenstates of $\sigma_x$ and $\sigma_z$. Then, we adopt the same (unital, trace-preserving) channel $\Lambda$ as specified in the main text, with the same coefficients as used to robustly self-test the preparations: $c(\theta)=\min\{1,\frac{s}{4}\sin\theta\}$ when $\theta\in [0,\pi/4]$ and $c(\theta)=\min\{1,\frac{s}{4}\cos\theta\}$ when $\theta\in(\pi/4,\pi/2]$.

It is straightforward to see that, for any given pair of measurements, the optimal choice of preparations are four pure qubit states, such that $\rho_{00}$ and $\rho_{11}$, and $\rho_{01}$ and $\rho_{10}$ respectively, correspond to antipodal vectors on the Bloch sphere. Therefore, we can without loss of generality restrict to such preparations since these impose the weakest constraints on the measurements of our interest. We can therefore parametrise the preparations $\rho_{x_0x_1}=1/2\left(\openone+\vec{m}_{x_0x_1}\cdot \vec{\sigma}\right)$ by Bloch vectors
\begin{align}
& \vec{m}_{00}=\left[\cos\theta,0,\sin\theta\right] & \vec{m}_{11}=-\left[\cos\theta,0,\sin\theta\right] && \vec{m}_{01}=\left[\cos\theta,0,-\sin\theta\right] && \vec{m}_{10}=\left[-\cos\theta,0,\sin\theta\right].
\end{align}
Expressing $Z_{yb}$ in terms of these preparations gives
\begin{align}
& Z_{00}=\frac{1}{8}\left(\openone+\cos\theta \sigma_x\right) & Z_{01}=\frac{1}{8}\left(\openone-\cos\theta \sigma_x\right) 
&& Z_{10}=\frac{1}{8}\left(\openone+\sin\theta \sigma_z\right) \sigma_z && Z_{11}=\frac{1}{8}\left(\openone-\sin\theta \sigma_z\right).
\end{align}
Due to symmetries, we restrict ourselves so that $t_o\equiv t_{01}=t_{10}$ and $t_e\equiv t_{00}=t_{11}$. Thus, we have to consider two operator inequalities in each interval $\theta\in [0,\pi/4]$ and $\theta\in (\pi/4,\pi/2]$. In the first interval, the two operator inequalities are
\begin{align}
& \frac{1+\sigma_x}{2}-\frac{s}{8}\left(\openone+\cos\theta \sigma_x\right) -t_e\openone\geq 0 &
& \frac{1+c(\theta)\sigma_z}{2}-\frac{s}{8}\left(\openone+\sin\theta \sigma_z\right)-t_o\openone\geq 0.
\end{align}
In the second interval, the two operator inequalities are
\begin{align}
& \frac{1+c(\theta)\sigma_x}{2}-\frac{s}{8}\left(\openone+\cos\theta \sigma_x\right)-t_e\openone\geq 0 &
& \frac{1+\sigma_z}{2}-\frac{s}{8}\left(\openone+\sin\theta \sigma_z\right)-t_o\openone\geq 0.
\end{align}
Just as in Appendix \ref{B}, we solve these inequalities for $t_e$ and $t_o$, and choose the largest value compatible with the solutions. In the first interval, this gives
\begin{align}
& t_e=\min\left\{\frac{1}{8}\left(8-s-s\cos\theta\right) ,\frac{s}{8}\left(\cos\theta-1\right)\right\} & t_o=\min \left\{\frac{1}{8} (4 c(\theta)-s \sin \theta-s+4),\frac{1}{8} (-4 c(\theta)+s \sin \theta-s+4)\right\}.
\end{align}
A similar procedure for the interval $\theta\in (\pi/4,\pi/2]$ leads to 
\begin{align}
& t_e=\min \left\{\frac{1}{8} (4 c(\theta)-s \cos \theta-s+4),\frac{1}{8} (-4 c(\theta)+s \cos \theta-s+4)\right\} & t_o=\min\left\{\frac{s}{8} (\sin \theta -1) ,\frac{1}{8} (8-s -s\sin \theta)\right\}.
\end{align}
For any choice of $\theta$, we have constructed operator inequalities of the form \eqref{opm}. 

In order to obtain our lower bound on $\mathcal{F}'$, we must minimise the quantity $t(\theta)=(t_e+t_o)/2$ for a specific choice of $s$. In analogy with the procedure in Appendix \ref{C}, we choose $s=4\left(1+\sqrt{2}\right)$, which returns $\min_\theta t(\theta)=-3/(2\sqrt{2})$. Hence, we have obtained the lower bound  
\begin{equation}
\mathcal{F}'\left(\mathcal{A}_2\right)\geq \left(1+\sqrt{2}\right)\mathcal{A}_2-\frac{3}{2\sqrt{2}} = L(\mathcal{A}_2) .
\end{equation}

\section{Self-testing all pairs of incompatible Pauli observables}\label{Cbisbis}

Consider a generalisation of the $2\rightarrow 1$ RAC, in which we introduce a bias on the score associated to certain inputs. Specifically, whenever the game is successful, i.e. $b=x_y$, the awarded score is $q/2$ if $x_0 \oplus x_1=0$, and $(1-q)/2$ if $x_0 \oplus x_1=1$, for some $q\in [0,1]$. The average score reads
\begin{equation}
\mathcal{A}_{2}^q=\frac{1}{2}\sum_{x_0,x_1,y}r(x_0,x_1)P(b=x_y|x_0,x_1,y),
\end{equation} 
where $r(x_0,x_1)=q/2$ if $x_0 \oplus x_1=0$ and $r(x_0,x_1)=(1-q)/2$ if $x_0 \oplus x_1=1$. Note that for $q=1/2$, we recover the standard $2\rightarrow 1$ RAC. Based on the quantity $\mathcal{A}_{2}^q$, we will now see how to derive a self-testing condition for any pair of incompatible Pauli observables, i.e. any pair of non-commuting projective rank-one qubit measurements.

We start by expressing $\mathcal{A}_{2}^q$ for a quantum strategy:
\begin{multline}
\mathcal{A}_{2}^q=\frac{1}{2}+\frac{1}{4}\sum_{x_0,x_1}r(x_0,x_1)\Tr\left(\rho_{x_0x_1}\left((-1)^{x_0}M_0+(-1)^{x_1}M_1\right)\right)\leq \frac{1}{2}+\frac{1}{4}\sum_{x_0,x_1}r(x_0,x_1)\lambda_{\text{max}}\left[(-1)^{x_0}M_0+(-1)^{x_1}M_1\right]. 
\end{multline}
Denoting $\mu_k=\lambda_{\text{min}}\left[M_0+(-1)^{k}M_1\right]$ and $\nu_k=\lambda_{\text{max}}\left[M_0+(-1)^{k}M_1\right]$, for $k=0,1$, we obtain
\begin{equation}
\mathcal{A}_2^q\leq \frac{1}{2}+\frac{1}{8}\left[q\left(\mu_0-\nu_0\right)+(1-q)\left(\mu_1-\nu_1\right)\right].
\end{equation} 
Following a derivation analogous to that appearing in Appendix \ref{A} to obtain, we obtain
\begin{equation}\label{biasedbound}
\mathcal{A}_2^q\leq \frac{1}{2}+\frac{1}{8}\left[q\sqrt{\beta+\alpha}+(1-q)\sqrt{\beta-\alpha}\right],
\end{equation}
where $\beta=2\Tr\left(M_0^2+M_1^2\right)-\Tr\left(M_0\right)^2-\Tr\left(M_1\right)^2$ and $\alpha=2\Tr\left(\{M_0,M_1\}\right)-2\Tr\left(M_0\right)\Tr\left(M_1\right)$. Treating $\alpha$ and $\beta$ as independent variables, we obtain the largest value of the right-hand-side of Eq.~\eqref{biasedbound} by demanding that the derivative with respect to $\alpha$ equals zero, and checking that the second derivative is negative at this point. We obtain the optimality constraint
\begin{equation}\label{cons}
\alpha=\frac{2q-1}{1-2q+2q^2}\beta.
\end{equation}
Inserting this value back into Eq.~\eqref{biasedbound}, we find an upper bound on $\mathcal{A}_2^q$ as obtained by independent variables $\alpha$ and $\beta$. It turns out that this bound can be saturated by the de facto coupled variables $\alpha$ and $\beta$. From Eq.~\eqref{biasedbound}, it is clear that a necessary condition for optimality is to maximise $\beta$. This amounts to the observables $M_0$ and $M_1$ being traceless and such that $M_0^2=M_1^2=\openone$, leading to $\beta=8$. This implies that the observables represent projective rank-one measurements. Hence, we can write $M_y=\vec{n}_y\cdot \vec{\sigma}$ where the Bloch vector satisfies $|\vec{n}_y|=1$. Hence, we have $\alpha=8\vec{n}_0\cdot \vec{n}_1$. Thus, Eq.\eqref{cons} becomes
\begin{equation}
\vec{n}_0\cdot\vec{n}_1=\frac{2q-1}{1-2q+2q^2},
\end{equation}
which has a solution for any choice of $q$. Note that setting $q=1/2$ reduces the above to  $\vec{n}_0\cdot\vec{n}_1=0$ which we recognise as the optimality constraint for the standard $2\rightarrow 1$ random access code. In conclusion, for any pair of incompatible Pauli observables (characterised by the scalar product $\vec{n}_0\cdot\vec{n}_1$), we have a game $\mathcal{A}_2^q$ (where $q$ is chosen in order to satisfy the above equation), such that the maximal score can only be attained by using that specific pair of Pauli observables. We thus obtain a general class of self-tests for any pair of Pauli observables, corresponding to saturating the maximal quantum value of $\mathcal{A}_2^q$ for a given value of $q$:
\begin{equation}
\mathcal{A}_2^q\leq \frac{1}{2}\left(1+\sqrt{1-2q+2q^2}\right).
\end{equation}

\section{Self testing for the $N\rightarrow 1$ random access code}\label{D}

In this appendix, we extend the results presented in the main text to self-test the preparations and measurements in an $N\rightarrow 1$ RAC. The latter is a straightforward generalisation of the $2\rightarrow 1$ RAC considered in the main text. The input of the preparation device is a random $N$-bit string $x\equiv (x_1,\ldots,x_N)$, while the input of the measurement device is $y\in\{1,\ldots,N\}$. The average score is
\begin{equation}
\mathcal{A}_N=\frac{1}{N2^N}\sum_{x,y}P(b=x_y|x,y).
\end{equation}
Considering qubit states $\rho_{x}$, and measurement observables $M_y$, we get
\begin{equation}\label{s1}
\mathcal{A}_N=\frac{1}{2}+\frac{1}{N2^{N+1}}\sum_{x,y}(-1)^{x_y}\Tr\left(\rho_xM_y\right).
\end{equation}

\subsection{Compatibility of measurements}
We determine whether a set of measurements can explain (i.e. are compatible with) a given value of $\mathcal{A}_N$. Since rank-one projective measurements are optimal for any set of preparations, we choose for simplicity to restrict our consideration to such measurements. However, it is straightforward to consider general measurements using the method outlined in the main text and Appendix \ref{A}.

Specifically, we first write
\begin{equation}\label{racc}
\mathcal{A}_N=\frac{1}{2}+\frac{1}{N2^{N+1}}\sum_{x}\Tr\left(\rho_x W_x\right) \leq \frac{1}{2}+\frac{1}{N2^{N+1}}\sum_{x}\lmax\left[W_{x}\right],
\end{equation}
where $W_{x}=\sum_{y}(-1)^{x_y}M_y$.

Note $\lmax\left[W_{x}\right] = \lambda_{\text{min}}\left[W_{\bar{x}}\right]$, where $\bar{x} = (\bar{x}_1,...,\bar{x}_N)$ the bit string obtained from $x$ by flipping all bits. Thus it is sufficient to only calculate eigenvalues for the strings not obtainable from each other under a full bit-flip operation. To this end let $z=x_1\ldots x_{N-1},0$ and $\lambda_{z,0}$ ($\lambda_{z,1}$) be the largest (smallest) eigenvalue of $W_z$. Thus we write
\begin{equation}\label{racN}
\mathcal{A}_N\leq \frac{1}{2}+\frac{1}{N2^{N+1}}\sum_{z}\left[\lambda_{z,0}-\lambda_{z,1}\right].
\end{equation}
Since $\lambda_{z,0}^2$ and  $\lambda_{z,1}^2$ are eigenvalues of $W_z^2$, we have $\lambda_{z,0}^2+\lambda_{z,1}^2=\Tr\left(W_z^2\right)$, which is equivalent to 
\begin{equation}
\lambda_{z,0}^2+\lambda_{z,1}^2=\sum_{y=1}^{N}\Tr\left(M_y^2\right)+\sum_{k<l} (-1)^{z_k+z_l}\Tr\left(\{M_k,M_l\}\right).
\end{equation}
This equation, together with the relation $\left(\lambda_{z,0}-\lambda_{z,1}\right)^2\leq 2\left(\lambda_{z,0}^2+\lambda_{z,1}^2\right)$, imply that Eq.~\eqref{racN} becomes 
\begin{equation}\label{mbound}
\mathcal{A}_N\leq \frac{1}{2}+\frac{\sqrt{2}}{N2^{N+1}}\sum_{z}\Bigg[\sum_{y=1}^{N}\Tr\left(M_y^2\right)+\sum_{k<l} (-1)^{z_k+z_l}\Tr\left(\{M_k,M_l\}\right)\Bigg]^{1/2}.
\end{equation}

This provides a robust self-testing condition, allowing one to determine whether a given set of measurements is compatible with the observed value of $\mathcal{A}_N$. Furthermore, we can derive an upper bound on the maximal value of $\mathcal{A}_N$ by assuming (incorrectly for $N>3$) that there exists $N$ mutually unbiased bases in $\mathbb{C}^2$. This means that all measurements are maximally incompatible, i.e. that $\Tr\left(\{M_k,M_l\}\right)=0$ for $k\neq l$. Consequently, Eq.~\eqref{mbound} reduces to
\begin{equation}
\mathcal{A}_N\leq \frac{1}{2}\left(1+\frac{1}{\sqrt{N}}\right).
\end{equation}
We emphasise that only three mutually unbiased bases exist in $\mathbb{C}^2$ and hence this bound is only tight for $N=2,3$. For $N=2$, we recover the result presented in the main text. For $N=3$, this implies that a maximal value of $\mathcal{A}_3$ (i.e.~achieving the right-hand side of the above inequality) ensures that the measurements are three mutually unbiased qubit observables, such as the three Pauli observables $\sigma_x$, $\sigma_y$ and $\sigma_z$.

Going one step further, we can then also self-test the preparations (still assuming maximal value of $\mathcal{A}_3$). Indeed, each preparation $\rho_x$ must be pure, and correspond to the eigenvector of $W_x$ associated to its largest eigenvalue. Such a set of preparations correspond to a set of Bloch vectors forming a cube on the surface of the Bloch sphere. 

\subsection{Compatibility of preparations}

We ask whether a given value of $\mathcal{A}_N$ can be explained by a particular set of preparations. We suitably express \eqref{s1} in a quantum model and subsequently apply the Cauchy-Schwarz inequality for operators to obtain
\begin{equation}\label{prep2}
\mathcal{A}_N=\frac{1}{2}+\frac{1}{N2^N}\sum_{y=1}^{N}\Tr\left[M_y^0\sum_{x} (-1)^{x_y}\rho_{x}\right]
\leq \frac{1}{2}+\frac{1}{N2^N}\sum_{y=1}^{N}\sqrt{\Tr\left[M_y^0\left(\sum_{x} (-1)^{x_y}\rho_{x}\right)^2\right]}.
\end{equation}
In the last expression, the squared operator is evaluated to
\begin{equation}\label{op2}
\left(\sum_{x} (-1)^{x_y}\rho_x\right)^2=\sum_{x}\rho_x^2+\sum_{k<l}(-1)^{k_y+l_y}\{\rho_{k},\rho_{l}\}.
\end{equation}
If necessary, the anticommutators can be evaluated using Bloch sphere representation with the relation $\{\rho_k,\rho_l\}=1/2\left( (1+\vec{m}_k\cdot \vec{m}_l)\openone+(\vec{m}_k+\vec{m}_l)\cdot\vec{\sigma}\right)$. However, it is more convenient to consider a basis-independent representation. Importantly, note that since an equal number of positive and negative terms appear inside the square, the operator $\sum_{x} (-1)^{x_y}\rho_x$ is a linear combination of $\{\sigma_x,\sigma_y,\sigma_z\}$ and hence its square is proportional to the identity operator. Therefore, when re-inserting Eq.~\eqref{op2} into Eq.~\eqref{prep2}, we find
\begin{equation}\label{nself2}
\mathcal{A}_N\leq \frac{1}{2}+\frac{1}{N2^N} \sum_{y=1}^{N} \left[\sum_{x}\Tr\left(\rho_x^2\right)+\sum_{k<l}(-1)^{k_y+l_y}\Tr\left(\{\rho_{k},\rho_{l}\}\right)\right]^{1/2}.
\end{equation}
This is a self-testing condition for preparations, assessing whether a given set of preparations is compatible with a given value of $\mathcal{A}_N$.  In particular, a classical strategy in which the preparations are binary messages corresponds to $\forall x: \Tr\left(\rho_x^2\right)=1$ and  $\Tr\left(\{\rho_{k},\rho_{l}\}\right)=2\delta_{E(k),E(l)}$, where $E$ is the specific classical encoding strategy, i.e. a function $E:\{0,1\}^N\rightarrow \{0,1\}$.

\section{Self-testing with three-level systems}\label{F}
In the main text, we have considered self-testing in the $2 \rightarrow 1$ random access code when the physical system transmitted from Alice to Bob is a qubit. Clearly, if that system is allowed to carry two bits of information, the task is trivial since Alice can send both her inputs to Bob. Here, we consider the remaining non-trivial case of Alice communicating a three-level quantum system. To simplify the analysis we restrict ourselves to projective measurements for which all possible arrangements admit a compact characterisation. We show that the optimal quantum value equals $\mathcal{A}_{2} = \left(5+\sqrt{5}\right) / 8 \approx 0.9045$ and find all the optimal arrangements of observables (we argue that the optimal value is achieved only if both measurements are projective). Our argument is robust in the sense that we are able to certify incompatibility of $M_{0}$ and $M_{1}$ whenever the success probability exceeds the classical bound for three-level systems, which turns out to be $\mathcal{A}_{2} \leq 7/8$.

To obtain a statement which only depends on the observables we follow the main text and evaluate the sum
\begin{equation}
\sum_{x_{0}, x_{1}} \lmax \big[ (-1)^{x_0} M_{0}+(-1)^{x_1} M_{1} \big].
\end{equation}
Jordan's lemma states that any two projective observables can be simultaneously diagonalised such that the resulting blocks are $1 \times 1$ or $2 \times 2$. For observables acting on a qutrit, we only need to consider two cases: (a) three 1-dimensional subspaces or (b) one subspace of each type. Case (a) corresponds to classical strategies and it is easy to check that these satisfy $\mathcal{A}_{2} \leq 7/8$. In case (b) the observables (up to a unitary) can be written as
\begin{equation}
M_{0} = \left(
\begin{array}{cc}
\cos \alpha \, \sigma_{x} + \sin \alpha \, \sigma_{z} &\\
& r
\end{array}
\right)
\quad \textnormal{and} \quad
M_{1} = \left(
\begin{array}{cc}
\cos \alpha \, \sigma_{x} - \sin \alpha \, \sigma_{z} &\\
& s
\end{array}
\right)
\end{equation}
for some angle $\alpha \in [0, 2 \pi]$ and $r, s \in \{ \pm 1\}$. A simple calculation yields
\begin{align*}
&\lmax[ M_{0} + M_{1} ] = \max \{ 2 \left| \cos \alpha \right|, r + s \},\\
&\lmax[ M_{0} - M_{1} ] = \max \{ 2 \left| \sin \alpha \right|, r - s \},\\
&\lmax[ - M_{0} + M_{1} ] = \max \{ 2 \left| \sin \alpha \right|, - r + s \},\\
&\lmax[ - M_{0} - M_{1} ] = \max \{ 2 \left| \cos \alpha \right|, - r - s \}
\end{align*}
and, therefore,
\begin{equation}
\label{eq:three-level-bound}
\sum_{x_{0}, x_{1}} \lmax \big[ (-1)^{x_0} M_{0}+(-1)^{x_1} M_{1} \big] =
\begin{cases}
2 + 4 \left| \sin \alpha \right| + 2 \left| \cos \alpha \right| \quad \textnormal{if} \quad r = s,\\
2 + 2 \left| \sin \alpha \right| + 4 \left| \cos \alpha \right| \quad \textnormal{if} \quad r \neq s.
\end{cases}
\end{equation}
For $r = s$ the right-hand side is maximised for $\alpha \in \{ c_{1}, c_{1} + \pi, - c_{1} + \pi, - c_{1} + 2 \pi  \}$, where $c_{1}$ is the unique solution to $\tan c_{1} = 2$ in the interval $[0, \pi/2]$. Similarly, for $r \neq s$ the right-hand side is maximised for $\alpha \in \{ c_{2}, c_{2} + \pi, - c_{2} + \pi, - c_{2} + 2 \pi  \}$, where $c_{2}$ is the unique solution to $\tan c_{2} = 1/2$ in the interval $[0, \pi/2]$.

While the different optimal arrangements are not unitarily equivalent, they are of similar form. The optimal arrangement characterised by $r = s = 1$ and $\alpha = c_{1}$ yields the following optimal preparations
\begin{equation}
\rho_{00} = \left(
\begin{array}{cc}
0 &\\
& 1
\end{array}
\right), \quad
\rho_{01} = \left(
\begin{array}{cc}
(\openone + \sigma_{z})/2 &\\
& 0
\end{array}
\right), \quad
\rho_{10} = \left(
\begin{array}{cc}
(\openone - \sigma_{z})/2 &\\
& 0
\end{array}
\right), \quad
\rho_{11} = \left(
\begin{array}{cc}
(\openone - \sigma_{x})/2 &\\
& 0
\end{array}
\right).
\end{equation}
Indeed, it is always the case that one preparation lives in the $1 \times 1$ subspace, whereas the other three occupy the $2 \times 2$ subspace (two of them form a basis to which the last one is unbiased). To see that the optimal winning probability requires projective measurements note that for every set of preparations the optimal observables can be chosen projective. However, all sets of preparations optimal for projective observables are of the form given above and one can check that for these preparations the optimal measurements must be projective (a direct consequence of the fact that the operators $ \rho_{00} + \rho_{01} - \rho_{10} - \rho_{11}$ and $ \rho_{00} - \rho_{01} + \rho_{10} - \rho_{11}$ are full-rank).

It is the presence of multiple inequivalent maximisers that prevents us from writing down a simple self-testing statement. However, Eq.~\eqref{eq:three-level-bound} allows us to deduce the range of $\alpha$ compatible with the observed value of $\mathcal{A}_{2}$ (note that the conclusion will be stronger if we know whether $r = s$ or $r \neq s$). In particular, any value exceeding the classical bound of $7/8$ implies a lower bound on the incompatibility between $M_{0}$ and $M_{1}$ on the $2 \times 2$ subspace.


	\section{Numerical method for robust self-testing}
	In the main text, we focused on the RAC and derived an optimal robust self-test. However, robust self-testing is relevant also for many other tasks that are not RACs. Here, we outline a numerical method based on semi-definite programming for inferring lower bounds on the worst-case average fidelity of preparations $\mathcal{F}$ in more general tasks. Specifically, we adapt the so-called swap-method of \cite{swaptrick} (constructed for Bell scenarios) to prepare-and-measure scenarios by combining it with the hierarchy of dimensionally bounded quantum correlations \cite{NV}. For sake of instruction, we first present the method by applying it to the RAC, and then use it to robustly self-test preparations in another prepare-and-measure scenario.
	
	The preparations in the random access code are self-tested up to a collective unitary transformation. A robust self-test must therefore be valid under this degree of freedom. However, one can only consider the fidelity of the unknown preparations with respect to the optimal states in some choosen basis. Therefore, in order to achieve a robust self-test, one needs to find a way to avoid the possibility of a collective unitary misaligning the bases.  This can be done by supplying Bob's measurement device with an auxillary system, say it is initialised in the state $\ket{0}_{\text{A}}$, into which the unknown received preparations can be swaped \cite{swaptrick}. In the RAC, the optimal measurements are anticommuting Pauli measurements. Therefore, with inspiration from this ideal case, Bob's swap operator $S$ can be composed as follows: $S=UVU$, where
	\begin{align}\label{swap}
	&U=\openone \otimes \ket{0}\bra{0}+B_1\otimes\ket{1}\bra{1} & V=\frac{1+B_0}{2}\otimes \openone+\frac{1-B_0}{2}\otimes \sigma_x,
	\end{align}
	where $B_0$ and $B_1$ denote the observables of Bob. If $B_0$ and $B_1$ correspond to $\sigma_z$ and $\sigma_x$ respectively, the above returns the two-qubit swap operator. Bob applies $S$ to the joint system of received preparation (labeled B) and ancilla (labeled A). The state swaped into Bob's ancilla reads
	\begin{equation}
	\rho^{\textnormal{SWAP}}_{x_0x_1}=\Tr_{\text{B}}\left[S\left(\rho_{x_0x_1}\otimes \ket{0}_{\text{A}} {_\text{A}}\bra{0}\right)S^\dagger\right].
	\end{equation}
	Consequently, the worst-case average fidelity of Alice's preparations with her optimal preparations is
	\begin{equation}\label{fid2}
	\mathcal{F}(\mathcal{A}_2^*)=\min_{\rho\in R(\mathcal{A}_2^*)}\max_{\Lambda}\frac{1}{4}\sum_{x_0x_1} \Tr\left[\Lambda[\rho^{\textnormal{ideal}}_{x_0x_1}]\rho^{\textnormal{SWAP}}_{x_0x_1}\right]=\min_{\rho\in R(\mathcal{A}_2^*)}\max_{\Lambda}\frac{1}{4}\sum_{x_0x_1} \Tr\left[S\left(\Lambda[\rho_{x_0x_1}]\otimes \ket{0}_{\text{A}}{_\text{A}}\bra{0}\right)S^\dagger \left(\openone \otimes \rho^{\textnormal{ideal}}_{x_0x_1}\right)\right],
	\end{equation}
	where $R(\mathcal{A}_2^*)$ is the set of all preparations that are compatible with the value $\mathcal{A}_2^*$, and $\Lambda$ is the extraction channel, the duality of which is used above.
	
	We may write the operator $S$ in terms Bob's observables as follows:
	\begin{equation}
	S=\frac{1}{2}\sum_{ij}s_{ij}\otimes \ket{i}_{\text{A}}{_\text{A}}\bra{j},
	\end{equation}
	where 
	\begin{equation}
	s_{00}=\openone+B_0 \hspace{9mm} s_{01}=B_1-B_0B_1 \hspace{9mm} s_{10}=B_1-B_1B_0 \hspace{9mm} s_{11}=\openone+B_1B_0B_1.
	\end{equation}
	Inserting this into \eqref{fid2} we find
	\begin{align}\label{fid3}\nonumber
	\mathcal{F}(\mathcal{A}_2^*)&=\min_{\rho\in R(\mathcal{A}^*)}\max_{\Lambda}\frac{1}{16}\sum_{x_0x_1} \sum_{ijkl} \Tr\left[\left(s_{ij}\otimes \ket{i}_{\text{A}}{_\text{A}}\bra{j}\right)\left(\Lambda[\rho_{x_0x_1}]\otimes \ket{0}_{\text{A}}{_\text{A}}\bra{0}\right)\left(s_{kl}\otimes\ket{k}_{\text{A}}{_\text{A}}\bra{l}\right)^\dagger\left(\openone\otimes \rho^{\textnormal{ideal}}_{x_0x_1}\right)\right]\\\nonumber
	&=\min_{\rho\in R(\mathcal{A}_2^*)}\max_{\Lambda}\frac{1}{16}\sum_{x_0x_1} \sum_{ijkl} \Tr\left[s_{ij}\Lambda[\rho_{x_0x_1}] s_{kl}^\dagger\right]\Tr\left[\ket{i}\braket{j}{0}\braket{0}{l}\bra{k}\rho^{\textnormal{ideal}}_{x_0x_1}\right]\\\nonumber
	&=\min_{\rho\in R(\mathcal{A}_2^*)}\max_{\Lambda}\frac{1}{16}\sum_{x_0x_1} \sum_{ik} \Tr\left[s_{k0}^\dagger s_{i0}\Lambda[\rho_{x_0x_1}]\right]\bra{k}\rho^{\textnormal{ideal}}_{x_0x_1}\ket{i}\\
	&=\min_{\rho\in R(\mathcal{A}_2^*)}\max_{\Lambda}\frac{1}{16}\sum_{x_0x_1}\sum_{ik}\Tr\left[T_{ik}\Lambda[\rho_{x_0x_1}]\right]\bra{k}\rho^{\textnormal{ideal}}_{x_0x_1}\ket{i},
	\end{align}
	where we defined $T_{ik}=s_{k0}^\dagger s_{i0}$. The four elements of $T$ are straightforwardly computed to
	\begin{align}
	&T_{00}=2\left(\openone+B_0\right), \hspace{9mm} T_{01}=B_1\left(\openone-B_0\right)+B_0B_1\left(\openone-B_0\right), \\
	&T_{11}=2\left(\openone-B_0\right),  \hspace{9mm} T_{10}=B_1\left(\openone+B_0\right)-B_0B_1\left(\openone+B_0\right) .
	\end{align}
	
	\begin{figure}
		\centering
		\includegraphics[width=0.5\columnwidth]{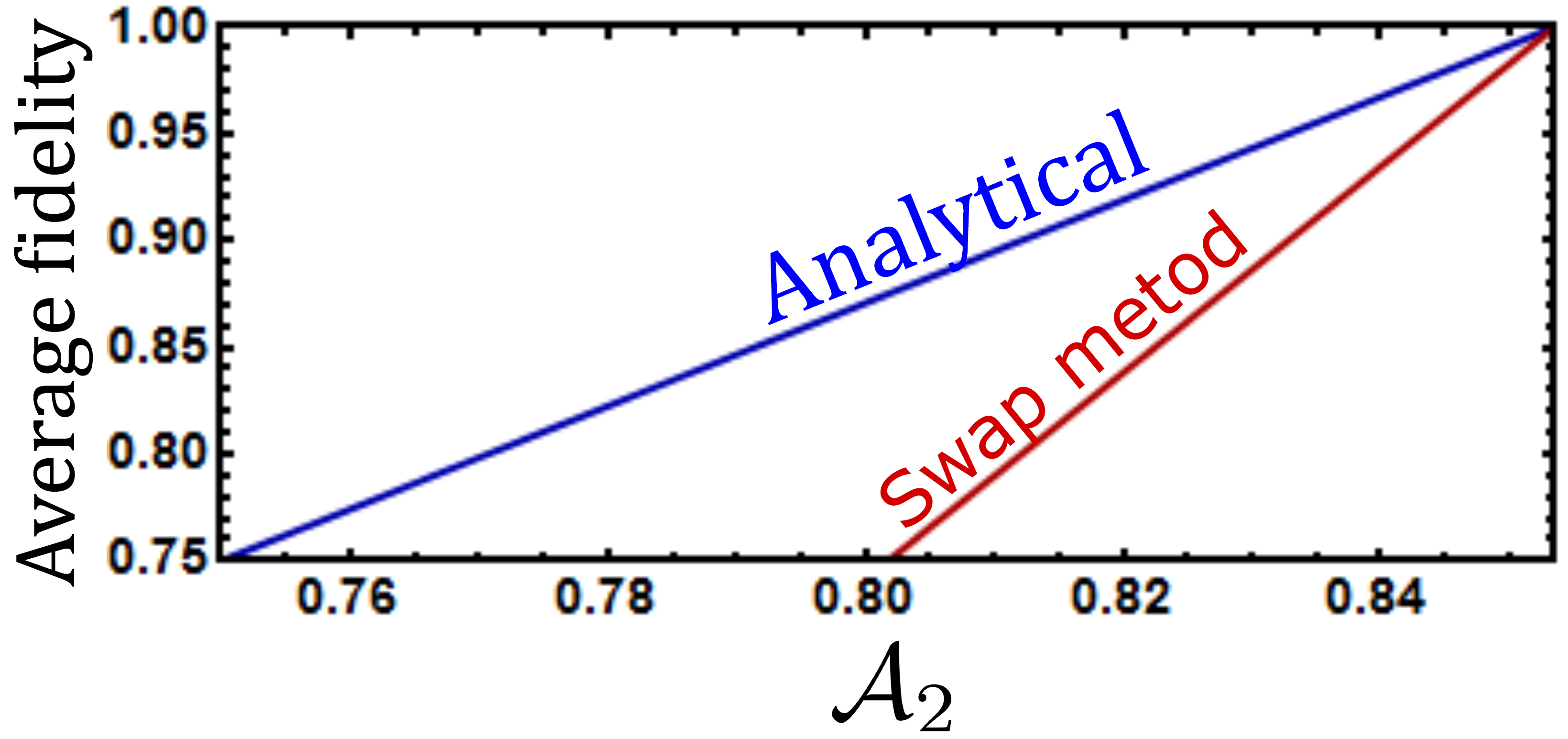}
		\caption{Lower bound on $\mathcal{F}(\mathcal{A}_2)$ as obtained by the swap method and by analytical technique.}\label{fig2}
	\end{figure}
	
	In the calculation of the fidelity, the same channel is applied to all Alice's preparations. We may simply consider that as four other valid preparations $\bar{\rho}_{x_0x_1}=\Lambda[\rho_{x_0x_1}]$. The fidelity in \eqref{fid3} is then a linear combination of variables $\{\Tr\left(\bar{\rho}_{x_0x_1}\openone\right),\Tr\left(\bar{\rho}_{x_0x_1}B_0\right),\ldots,\Tr\left(\bar{\rho}_{x_0x_1}B_0B_1B_0\right)\}$. Therefore, we may establish a lower bound on \eqref{fid3} using the dimensionally bounded hierarchy of quantum correlations \cite{NV}. The accuracy of this bound depends on the level of the hierarchy employed. We choose to consider the following level: we define a moment matrix 
	\begin{align}
	\chi_{ijkl}=\Tr\left[R^\dagger_j Q_i^\dagger Q_kR_l\right] \hspace{4mm} \text{where} \hspace{4mm }Q=\left(\openone, B_0, B_1, B_0B_1, B_1B_0\right) \hspace{4mm} \text{and} \hspace{4mm} R=\left(\openone, \bar{\rho}_{00}, \bar{\rho}_{01},\bar{\rho}_{10},\bar{\rho}_{11}\right),
	\end{align}
	for $i,j,k,l=1,\ldots,5$. From the moment matrix we calculate all terms needed to evaluate the average fidelity \eqref{fid3}, using the labels $x=2x_0+x_1+2$,
	\begin{align}
	&\Tr\left[T_{00} \bar{\rho}_{x_0x_1}\right]=2\chi_{111x}+2\chi_{112x} &\Tr\left[T_{11} \bar{\rho}_{x_0x_1}\right]=2\chi_{111x}-2\chi_{112x} \\
	&\Tr\left[T_{01} \bar{\rho}_{x_0x_1}\right]=\chi_{113x}+\chi_{114x}-\chi_{115x}-\chi_{215x} & \Tr\left[T_{10} \bar{\rho}_{x_0x_1}\right]=\chi_{113x}-\chi_{114x}+\chi_{115x}-\chi_{215x} .
	\end{align}
	In order to enforce that the average fidelity is extremised for a particular value $\mathcal{A}_2^*$ of the random access code, we write the probability distribution of Bob's outcomes in terms of the moment matrix as 
	\begin{equation}
	P(b\lvert x_0,x_1,y)=\frac{1+(-1)^b\chi_{1,1,y+2,x}}{2}.
	\end{equation}
	Thus, we can evaluate $\mathcal{A}_2$ as a linear combination of moment matrix elements. Fixing the value of  $\mathcal{A}_2$ corresponds to introducing an affine constraint on the moment matrix. Therefore, the following semi-definite program establishes a lower bound on  $\mathcal{F}(\mathcal{A}_2)$:
	\begin{align}\label{sdp}
	\mathcal{F}(\mathcal{A}_2^*)\geq \min_{\chi} \frac{1}{16}\sum_{x_0x_1}\sum_{i,k=0}^{1} \Tr\left(T_{ik}\bar{\rho}_{x_0x_1}\right)\langle k\lvert \rho^{\textnormal{ideal}}_{x_0x_1}\lvert i\rangle \\\nonumber
	\text{such that } \hspace{3mm}\chi\geq 0 \hspace{2mm} \text{  and  } \hspace{2mm} \mathcal{A}_2\geq \mathcal{A}_2^*.
	\end{align}
	
	We have implemented the semi-definite program and the results are presented in Fig.~\ref{fig2}, together with the lower bound on $\mathcal{F}(\mathcal{A}_2)$ obtained from the analytical method presented in the main text. Evidently, the swap method returns a sub-optimal but still non-trivial result. Using the swap method, we find a higher-than-classical value of $\mathcal{F}(\mathcal{A}_2)$, i.e., $\mathcal{F}(\mathcal{A}_2)>3/4$,  whenever $\mathcal{A}_2>0.802$.

	The advantage of the swap method is that it applies also to other prepare-and-measure scenarios beyond RACs. The drawback of the method is that the self-tests are typically not optimal, and that the complexity of evaluating the dimensionally bounded hierarchy of quantum correlations increases exponentially with the number of preparations and measurements, thus making more complicated scenarios infeasible to study.

\begin{figure}
		\centering
		\includegraphics[width=0.5\columnwidth]{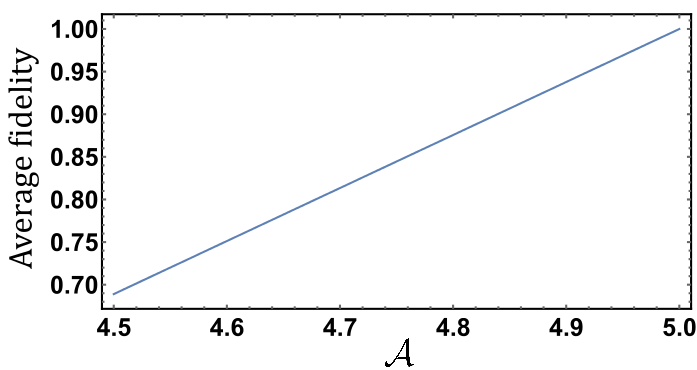}
		\caption{Lower bound on $\mathcal{F}(\mathcal{A})$ as obtained by the swap method.}\label{fig3}
	\end{figure}

	To examplify the usefulness of this method also for other prepare-and-measure scenarios, we present  a second example. Consider a prepare-and-measure scenario in which Alice has a random input $x\in\{0,1,2\}$ and Bob has a random input $y\in\{0,1\}$. Alice may only communicate a qubit to Bob. The objective of the scenario reads
	\begin{equation}
	\mathcal{A}=\sum_{x,y}c_{x,y}E(x,y),
	\end{equation}
	where $E(x,y)=p(b=0\lvert x,y)-p(b=1\lvert x,y)$ and $c_{x,0}=[1,1,-1]$ and $c_{x,1}=[\sqrt{3},-\sqrt{3},0]$. One straightforwardly finds that the maximal classical value  is $\mathcal{A}=1+2\sqrt{3}$. We wish to robustly self-test Alice's preparations solely based on the value of $\mathcal{A}$. From numerical brute-force maximisations of $\mathcal{A}$, we find that its maximal value is $\mathcal{A}=5$ and that this value is saturated using anticommuting Pauli measurements and preparations forming an equilateral triangle in a disk of the Bloch sphere. Such preparations can up to a unitary be written
	\begin{align}
	\rho_0^{\text{ideal}}=\frac{1}{2}\left(\openone+\sigma_x\right) && \rho_1^{\text{ideal}}=\frac{1}{2}\left(\openone+\frac{\sqrt{3}}{2}\sigma_z-\frac{1}{2}\sigma_x\right)  && \rho_2^{\text{ideal}}=\frac{1}{2}\left(\openone-\frac{\sqrt{3}}{2}\sigma_z-\frac{1}{2}\sigma_x\right). 
	\end{align}
	We make the ansatz that this constitutes a self-test of the preparations. We supply Bob with an ancilla state and define the swap operator as done in the RAC. Performing calculations fully analogous to the case of the RAC, we obtain a semi-definite program that gives a lower bound on the worst-case average fidelity 
	\begin{equation}
	\mathcal{F}(\mathcal{A})=\min_{\rho\in R(\mathcal{A})}\max_{\Lambda}\frac{1}{3}\sum_{x} \Tr\left[\Lambda[\rho^{\textnormal{ideal}}_{x}]\rho_{x}\right],
	\end{equation}
	where $\mathcal{R}(\mathcal{A})$ is the set of preparations compatible with the value $\mathcal{A}$ and $\Lambda$ is the extraction channel. We have used an intermediate level of the hierarchy of dimensionally bounded quantum correlations (sometimes referred to as 1+AB+BB+BBA) corresponding to an SDP matrix of size 20. The corresponding lower bound on $\mathcal{F}(\mathcal{A})$ is presented in Fig.~\ref{fig3}.
	We first see that that the maximal value $\mathcal{A}=5$ indeed self-tests (up to numerical precision) the preparations of Alice to form an equilateral triangle on the Bloch sphere (the fidelity is one). For non-maximal values of $\mathcal{A}$, we still obtain a non-trivial bound on the average fidelity of Alice's preparations with the optimal ones.


\end{document}